\documentclass[prc,showpacs,nofootinbib,preprint]{revtex4-1}
\usepackage{epsfig,graphics,color}
\usepackage{subfigure}
\usepackage{bbm}
\usepackage{mathtools}
\usepackage{amssymb}
\usepackage{mathrsfs}
\usepackage{amsmath}
\newcommand{\bra}[1]{\langle #1|}
\newcommand{\ket}[1]{|#1\rangle}

\newcommand{\roundbra}[1]{(#1|}
\newcommand{\roundket}[1]{|#1)}

\def\one{{\bf 1}\,}
\def\quarknumberoperator{{\mathbbm 1}\,}

\def\w2{\tilde w^2}
\def\ws2{1}

\usepackage{ulem}
\definecolor{mono}{rgb}{0.65, 0.65, 0.65} % Mono
\usepackage{pdfsync} % for sync via PDF reader w/ editor, but not necessary for compiling
\usepackage{dcolumn}
\newcolumntype{L}{>{$}l<{$}}
\newcolumntype{C}{>{$}c<{$}}
\newcolumntype{R}{>{$}r<{$}}
\begin{document}
\title{Large-$N_c$ sum rules for charmed baryons at subleading orders}
\author{Yonggoo Heo$^{1}$ and Matthias F.M. Lutz$^{2,3}$}
\affiliation{$^1$ Suranaree University of Technology, Nakhon Ratchasima, 30000, Thailand}
\affiliation{$^2$ GSI Helmholtzzentrum f\"ur Schwerionenforschung GmbH,\\
Planck Str. 1, 64291 Darmstadt, Germany}
\affiliation{$^3$ Technische Universit\"at Darmstadt, D-64289 Darmstadt, Germany}
\date{\today}
\begin{abstract}
Sum rules for the low-energy constants of the chiral SU(3) Lagrangian with charmed baryons of spin $J^P=1/2^+$ and $J^P=3/2^+$ baryons 
are derived from large-$N_c$ QCD. We consider the large-$N_c$ operator expansion at subleading orders for current-current correlation functions in the 
charmed baryon-ground states for two scalar and two axial-vector currents.
\end{abstract}

%%%%%%%%%%%%%%%%%%%%%%%%%%%%%%%%%%%%%%%
\maketitle

\section{Introduction}

The dependence of the charmed baryon masses on the up, down and strange quark masses encodes useful information on the coupled-channel 
interaction dynamics of the Goldstone bosons with such baryon states \cite{Lutz:2003jw,Hofmann:2005sw,Hofmann:2006qx,GarciaRecio:2008dp,JimenezTejero:2009vq,Lu:2014ina,Long:2015pua,Lu:2016gev}. Lattice QCD simulations for the baryon masses at unphysical 
quark masses are particularly useful \cite{Liu:2009jc,Bali:2012ua,Briceno:2012wt,Alexandrou:2012xk,Namekawa:2013vu,Perez-Rubio:2013oha} since they complement 
the well known values for the masses of the charmed baryon ground states at physical quark masses \cite{PDG}. 

An accurate flavour SU(3) chiral extrapolation of the baryon ground states with zero charm content was established in a series of works \cite{Semke2005,Semke2012,Semke:2012gs,Lutz:2014oxa,Lutz:2018cqo}. 
Based on the chiral Lagrangian formulated with spin 1/2 and 3/2 fields the available lattice data on the baryon masses was reproduced and accurate predictions for
the size of the low-energy parameters relevant at N$^3$LO were made \cite{Lutz:2018cqo}. 
The success of such analyses relies on two crucial ingredients. First, the chiral expansion is formulated in terms of physical meson and baryon masses rather than bare masses as is requested by traditional 
chiral perturbation theory ($\chi$PT). Second, the flood of low-energy constants that arises at subleading orders is tamed by sum rules for the latter as they arise in the limit of a large number of colors ($N_c$)
in QCD \cite{Lutz:2010se,Lutz:2018cqo}. The large-$N_c$ sum rules provide a large parameter reduction that allowed fits at N$^3$LO to the lattice
data set that are significant. A corresponding program was started for the charmed baryons \cite{Lutz:2014jja}. At present, however, the large-$N_c$ sum rules for the charmed baryons 
are derived at leading order only \cite{Lutz:2014jja}. It is the purpose of this work to close this gap and establish such sum rules accurate to subleading orders in the $1/N_c$ expansion. This will pave the 
way to accurate chiral extrapolation studies of the charmed baryon masses. 

The desired sum rules can systematically be derived from QCD by a study of current-current correlation functions in the baryon ground states. 
We study matrix elements of current-current correlation functions in the charmed baryon states \cite{Lutz:2010se,Lutz:2011fe}.
The technology developed in \cite{Luty1994,Dashen1994,Jenkins:1996de} will be applied.
The implications of heavy-quark symmetry on the counter terms was worked out already using a suitable multiplet
representation of the charmed baryons \cite{Yan:1992gz,Cho:1992gg,Casalbuoni:1996pg,Lutz:2014jja}.

\newpage

\section{Chiral dynamics for charmed baryons} \label{section:chiral-lagrangian}

The chiral dynamics for the charmed baryon fields is most economically deduced from an effective chiral Lagrangian that is based on power counting rules. 
We consider here the flavour antisymmetric anti-triplet and the flavour symmetric sextet fields $B_{[\bar 3]}$, 
$B_{[6]}$ and $B^\mu_{[6]}$  with $J^P = \frac{1}{2}^+$ and $J^P = \frac{3}{2}^+$ quantum numbers. The chiral Lagrangian 
consists of all possible interaction terms, formed with the baryon fields and the conventional chiral blocks $U_\mu $ and $\chi_\pm$ that include the 
Goldstone boson fields $\Phi$ as well as the classical source functions, $s, p $  and $ v_\mu, a_\mu $ of QCD \cite{Gasser:1984gg}. 
Derivatives of the fields must be included in compliance with the local chiral SU(3) symmetry which in turn requests the covariant
derivative $D_\mu$  to act on the various flavour multiplet fields as follows
\begin{eqnarray}
&&(D_\mu \, U_\nu)^{a}_{\;b} \;\,= \partial_\mu U^{a}_{\nu,b} +  \Gamma^{a}_{\mu,l}\, U^{l}_{\nu,b} -
\Gamma^{l}_{\mu,b}\, U^{a}_{\nu,l} \,, \quad
\nonumber\\
&&(D_\mu  B_{[6]})^{ab} = \partial_\mu B_{[6]}^{ab} +  \Gamma^{a}_{\mu,l}\, B^{lb}_{[6]} +
\Gamma^{b}_{\mu,l}\, B^{al}_{[6]} \,, \quad
\nonumber\\
&&(D_\mu  B_{[\bar 3]})^{ab} = \partial_\mu B_{[\bar 3]}^{ab} +  \Gamma^{a}_{\mu,l}\, B_{[\bar 3]}^{lb} +
\Gamma^{b}_{\mu,l}\, B_{[\bar 3]}^{al} \,,
\label{def-covariant-derivative}
\end{eqnarray}
with the chiral connection $\Gamma_\mu=-\Gamma_\mu^\dagger$  given by
\begin{eqnarray}
&&\Gamma_\mu ={\textstyle{1\over 2}}\,e^{-i\,\frac{\Phi}{2\,f}} \,\Big[\partial_\mu -i\,(v_\mu + a_\mu) \Big] \,e^{+i\,\frac{\Phi}{2\,f}}
+{\textstyle{1\over 2}}\, e^{+i\,\frac{\Phi}{2\,f}} \,
\Big[\partial_\mu -i\,(v_\mu - a_\mu)\Big] \,e^{-i\,\frac{\Phi}{2\,f}}\,,
\nonumber\\
&& \textcolor{black}{ U_\mu = {\textstyle{1\over 2}}\,u^\dagger \, \big(
\partial_\mu \,e^{i\,\frac{\Phi}{f}} \big)\, u^\dagger
-{\textstyle{i\over 2}}\,u^\dagger \,(v_\mu+ a_\mu)\, u
+{\textstyle{i\over 2}}\,u \,(v_\mu-a_\mu)\, u^\dagger\;, \qquad \qquad 
 u = e^{i\,\frac{\Phi}{2\,f}} } \,.
\end{eqnarray}
The various hadron fields can be decomposed into their isospin multiplet components
\begin{eqnarray}
&& \Phi = \tau \cdot \pi (140)
+ \alpha^\dagger \cdot  K (494) +  K^\dagger(494) \cdot \alpha
+ \eta(547)\,\lambda_8\,,
\nonumber\\
&&  \sqrt{2}\,B_{[\bar 3]}  = {\textstyle{1\over \sqrt{2}}}\,\alpha^\dagger \cdot \Xi_c(2470)
- {\textstyle{1\over \sqrt{2}}}\,\Xi_c^T(2470)\cdot \alpha
+  i\,\tau_2\,\Lambda_c(2284) \,,
\nonumber\\
&& \sqrt{2}\,B_{[6]} = {\textstyle{1\over \sqrt{2}}}\,\alpha^\dagger \cdot \Xi_c(2580)
+ {\textstyle{1\over \sqrt{2}}}\,\Xi_c^{T}(2580)\cdot \alpha
+ \Sigma_c(2455) \cdot \tau \,i\,\tau_2
\nonumber\\
&& \qquad \quad+  {\textstyle{\sqrt{2}\over 3}}\, \big(1-\sqrt{3}\,\lambda_8 \big)\,\Omega_c(2704)  \,,
\nonumber\\
&& \sqrt{2}\, B^\mu_{[6]} = {\textstyle{1\over \sqrt{2}}}\,\alpha^\dagger \cdot \Xi^{\mu}_c(2645)
+ {\textstyle{1\over \sqrt{2}}}\,\Xi_c^{T,\mu}(2645)\cdot \alpha
+ \Sigma^\mu_c(2520) \cdot \tau \,i\,\tau_2
\nonumber\\
&& \qquad \quad+  {\textstyle{\sqrt{2}\over 3}}\, \big(1-\sqrt{3}\,\lambda_8 \big)\,\Omega^{\mu}_c(2770)  \,,
\nonumber\\
&& \alpha^\dagger = {\textstyle{1\over \sqrt{2}}}\,(\lambda_4+i\,\lambda_5 ,\lambda_6+i\,\lambda_7 )\,,\qquad
\tau = (\lambda_1,\lambda_2, \lambda_3)\,,
\end{eqnarray}
where the matrices $\lambda_i$ are the standard Gell-Mann generators of the SU(3) algebra.

The main goal of this work is to derive correlations amongst the low-energy parameters of the chiral Lagrangian
as they follow from a $1/N_c$ expansion. For that purpose we consider QCD's axial-vector and scalar currents,
\begin{eqnarray}
&& A_\mu^{(a)}(x) = \bar \Psi (x)\,\gamma_\mu \,\gamma_5\,\frac{\lambda_a}{2}\,\Psi(x) \,, \qquad \qquad
S^{(a)}(x) = \bar \Psi (x)\,\frac{\lambda_a}{2}\,\Psi(x) \,,
\label{def-amu}
\end{eqnarray}
in baryon matrix elements, where we recall their definitions in terms of the
Heisenberg quark-field operators $\Psi(x)$. With $\lambda_a$ we
denote the Gell-Mann flavour matrices supplemented with a singlet matrix $\lambda_0 = \sqrt{2/3}\,\one $.
Given the chiral Lagrangian, it is well defined how to derive the contribution of a given term to such matrix elements. The classical matrices of source functions, $a_\mu$ and $s$,
enter the chiral Lagrangian via the building blocks 
\begin{eqnarray}
U_\mu =  \frac{i}{2\,f}\,\partial_\mu \,\Phi - i\,a_\mu + \cdots \,, \qquad 
\chi_+ = {\textstyle{1\over 2}}\, \big(
u \,\chi_0 \,u
+ u^\dagger \,\chi_0 \,u^\dagger \big) = 2\,B_0\, s + \cdots \,,
\label{result:Umu}
\end{eqnarray}
where for notational simplicity in the following we put $B_0 = 1/2$.

We recall all terms in the chiral Lagrangian that are relevant in a chiral extrapolation of the baryon masses at N$^3$LO. 
Altogether we recalled $34+16= 50$ distinct low-energy constants, which have to be correlated.
There are 16 symmetry breaking counter terms
\begin{eqnarray}
&& \mathscr{L}_{\chi}^{(4)} \!= c_{1,[\bar 3\bar 3]}\,{\rm tr}\,\big( \bar B_{[\bar 3]}\,B_{[\bar 3]}\big)\,{\rm tr}\,\big(\chi_+^2\big)
+ c_{2,[\bar 3\bar 3]}\,{\rm tr}\,\big( \bar B_{[\bar 3]}\,B_{[\bar 3]}\big)\,\big({\rm tr}\,\chi_+\big)^2
\nonumber\\
&& \qquad \;+\, c_{3,[\bar 3\bar 3]}\,{\rm tr}\,\big( \bar B_{[\bar 3]}\,\chi_+\,B_{[\bar 3]}\,\big)\,{\rm tr}\,\big(\chi_+\big)
+  c_{4,[\bar3\bar3]}\,{\rm tr}\,\big( \bar B_{[\bar3]}\,\chi_+^2\,B_{[\bar3]}\big)
\nonumber\\
&& \qquad \; + \,c_{1,[66]}\,{\rm tr}\,\big( \bar B_{[6]}\,B_{[6]}\big)\,{\rm tr}\,\big(\chi_+^2\big)
+ c_{2,[66]}\,{\rm tr}\,\big( \bar B_{[6]}\,B_{[6]}\big)\,\big({\rm tr}\,\chi_+\big)^2
\nonumber\\
&& \qquad \;+\, c_{3,[66]}\,{\rm tr}\,\big( \bar B_{[6]}\,\chi_+\,B_{[6]}\,\big)\,{\rm tr}\,\big(\chi_+\big)
+ c_{4,[66]}\,{\rm tr}\,\big( \bar B_{[6]}\,\chi_+^2\, B_{[6]}\big)
+ c_{5,[66]}\,{\rm tr}\,\big( \bar B_{[6]}\,\chi_+\, B_{[6]}\,\chi^T_+\big)
\nonumber\\
&& \qquad \;+\, c_{1,[\bar 3 6]}\,{\rm tr}\,\big( \bar B_{[6]}\,\chi_+\,B_{[\bar 3]}+ {\rm h.c.}\,\big)\,{\rm tr}\,\big(\chi_+\big)
+  c_{2,[\bar3 6]}\,{\rm tr}\,\big( \bar B_{[6]}\,\chi_+^2\,B_{[\bar3]}+ {\rm h.c.}\big)
\nonumber\\
%&& \qquad \;\textcolor{mono}{
%  +\, c_{3,[\bar3 6]}\,{\rm tr}\,\big( \bar B_{[6]}\,\chi_+\, B_{[\bar3]}\,\chi^T_+ + {\rm h.c.}\big)
%}
%\nonumber\\
&& \qquad \; -\, e_{1,[66]}\,{\rm tr}\,\big( \bar B_{[6]}^\mu\,g_{\mu \nu }\,B_{[6]}^\nu\big)\,{\rm tr}\,\big(\chi_+^2\big)
- e_{2,[66]}\,{\rm tr}\,\big( \bar B^\mu_{[6]}\,g_{\mu \nu}\,B_{[6]}^\nu\big)\,\big({\rm tr}\,\chi_+\big)^2
\nonumber\\
&& \qquad \;-\, e_{3,[66]}\,{\rm tr}\,\big( \bar B^\mu_{[6]}\,\chi_+\,g_{\mu \nu}\,B_{[6]}^\nu\,\big)\,{\rm tr}\,\big(\chi_+\big)
- e_{4,[66]}\,{\rm tr}\,\big( \bar B^\mu_{[6]}\,g_{\mu \nu}\,\chi_+^2\, B_{[6]}^\nu\big)
\nonumber\\
&& \qquad \;-\, e_{5,[66]}\,{\rm tr}\,\big( \bar B^\mu_{[6]}\,g_{\mu \nu}\,\chi_+\, B_{[6]}^\nu\,\chi^{\textcolor{black}{T}}_+\big)\,,
\label{def-L4}
\end{eqnarray}
which contribute to the baryon masses at tree-level. Not that as compared to \cite{Lutz:2014jja}
we dropped the flavour redundant term proportional to $c_{3,[\bar 3 6]}$.
The symmetry breaking counter terms contribute to the current-current correlation
function of two time-ordered scalar currents
\begin{eqnarray}
&& S^{ab} (q) = i\,\int d^4 x \,e^{-i\,q\cdot x}  \,{\mathcal T}\,S^{(a)} (x)\,S^{(b)}(0) \,,
\label{def:scalar-scalar}
\end{eqnarray}
in the baryon states. We consider singlet and octet
components with $a,b = 0, \cdots ,8$. 

In addition there is a class of 34 symmetry conserving two-body counter terms that contribute to the baryon masses at the one-loop level. 
Following \cite{Lutz2002a,Lutz:2010se,Lutz:2014jja} the symmetry conserving counter term are classified according to their Dirac structure.
\allowdisplaybreaks[1]
\begin{eqnarray}
&&\mathscr{L}^{(2)} = \mathscr{L}^{(2)}_\chi +\mathscr{L}^{(S)} + \mathscr{L}^{(V)} + \mathscr{L}^{(A)} + \mathscr{L}^{(T)}\,.
\label{def-L2}
\end{eqnarray}
A complete list relevant at second order was constructed in \cite{Lutz:2014jja} with
\begin{eqnarray}
&& \mathcal{L}^{(S)} \!= -g_{0,[\bar 3\bar 3]}^{(S)}\,{\rm tr}\,\big( \bar B_{[\bar 3]}\,B_{[\bar 3]}\big)\,{\rm tr}\,\big( U_\mu\,U^\mu\big)
% + reduandant g_{1,[\bar 3\bar 3]}^{(S)}\,{\rm tr}\,\big( \bar B_{[\bar 3]}\, U^\mu\,B_{[\bar 3]}\, U^T_\mu\big)
- g_{D,[\bar 3\bar 3]}^{(S)}\,{\rm tr}\,\big( \bar B_{[\bar 3]}\,\big\{ U_\mu,\,U^\mu\big\}\, B_{[\bar 3]}\big)
\nonumber\\
&& \qquad \;-\, 
g_{0,[66]}^{(S)}\,{\rm tr}\,\big( \bar B_{[6]}\,B_{[6]}\big)\,{\rm tr}\,\big( U_\mu\,U^\mu\big)
- g_{1,[66]}^{(S)}\,{\rm tr}\,\big( \bar B_{[6]}\, U^\mu\,B_{[6]}\, U^T_\mu\big)
\nonumber\\
&& \qquad \;-\, g_{D,[66]}^{(S)}\,{\rm tr}\,\big( \bar B_{[6]}\,\big\{ U_\mu,\,U^\mu\big\}\, B_{[6]}\big)
%\textcolor{mono}{- g_{1,[\bar 3 6]}^{(S)}\,{\rm tr}\,\big( \bar B_{[6]}\, U^\mu\,B_{[\bar 3]}\, U^T_\mu + {\rm h.c.}\big)}
-g_{D,[\bar 36]}^{(S)}\,{\rm tr}\,\big( \bar B_{[6]}\,\big\{ U_\mu,\,U^\mu\big\}\, B_{[\bar 3]} + {\rm h.c.}\big)
\nonumber\\
&& \qquad \;+\, h_{0,[66]}^{(S)}\,{\rm tr}\,\big( \bar B_{[6]}^\mu\,g_{\mu \nu}\,B_{[6]}^\nu\big)\,{\rm tr}\,\big( U_\alpha\,U^\alpha\big)
+ h_{1,[66]}^{(S)}\,{\rm tr}\,\big( \bar B_{[6]}^\mu\,B_{[6]}^\nu\big)\,{\rm tr}\,\big( U_\mu\,U_\nu\big)
\nonumber\\
&& \qquad \;+\, h_{2,[66]}^{(S)}\,{\rm tr}\,\big( \bar B_{[6]}^\mu\, g_{\mu \nu}\,\big\{ U^\alpha,\,U_\alpha\big\}\, B_{[6]}^\nu\big)
+ h_{3,[66]}^{(S)}\,{\rm tr}\,\big( \bar B_{[6]}^\mu\,\big\{ U_\mu,\,U_\nu\big\}\, B_{[6]}^\nu\big)
\nonumber\\
&& \qquad \;+\, h_{4,[66]}^{(S)}\,{\rm tr}\,\big( \bar B_{[6]}^\mu\,g_{\mu \nu}\,U^\alpha\, B_{[6]}^\nu\,U^T_\alpha \big)
+\frac{1}{2}\, h_{5,[66]}^{(S)}\,{\rm tr}\,\big( \bar B_{[6]}^\mu\,U_\nu\, B_{[6]}^\nu\,U^T_\mu + \bar B_{[6]}^\mu\,U_\mu\, B_{[6]}^\nu\,U^T_\nu\big)\,,
\nonumber\\
%%%%%%%%%%%%%%%%%%%%%%%%%%%%%%%%%%%%%%%%%
&& \mathcal{L}^{(V)} \!= -\frac{1}{2}\,g_{0,[\bar 3\bar 3]}^{(V)}\,{\rm tr}\,\big( \bar B_{[\bar 3]}\,i\,\gamma^\alpha\,(D^\beta B_{[\bar 3]})\,{\rm tr}\,\big( U_\beta\,U_\alpha\big) + {\rm h.c.} \big)
\nonumber\\
&& \qquad \;-\, \frac{1}{2}\,g_{1,[\bar 3\bar 3]}^{(V)}\,{\rm tr}\,\big( \bar B_{[\bar 3]}\,i\,\gamma^\alpha\,U_\beta\,(D^\beta B_{[\bar 3]})\, U^T_\alpha + \bar B_{[\bar 3]}\,i\,\gamma^\alpha\,U_\alpha\,(D^\beta B_{[\bar 3]})\, U^T_\beta + {\rm h.c.}\big)
\nonumber\\
&& \qquad \;-\,
 \frac{1}{2}\,g_{D,[\bar 3\bar 3]}^{(V)}\,{\rm tr}\,\big( \bar B_{[\bar 3]}\,i\,\gamma^\alpha\,\big\{ U_\alpha,\,U_\beta\big\}\,(D^\beta  B_{[\bar 3]}) + {\rm h.c.} \big)
\nonumber\\
%&& \qquad \;\textcolor{mono}{
%  -\,  \frac{1}{2}\,g_{1,[\bar 36]}^{(V)}\,{\rm tr}\,\big( \bar B_{[6]}\,i\,\gamma^\alpha\, U_\alpha\,(D^\beta  B_{[\bar 3]})\,U^T_\beta
%  + \bar B_{[6]}\,i\,\gamma^\alpha\, U_\beta\,(D^\beta  B_{[\bar 3]})\,U^T_\alpha  }
%  \nonumber\\ && \qquad\qquad
%  \textcolor{mono}{
%  - (D^\beta\bar{B}_{[6]})\,i\,\gamma^\alpha\,U_\alpha\,B_{[\bar{3}]}\,U^T_\beta
%  - (D^\beta\bar{B}_{[6]})\,i\,\gamma^\alpha\,U_\beta\,B_{[\bar{3}]}\,U^T_\alpha  + {\rm h.c.} \big)}
%\nonumber\\
&& \qquad \;-\,  \frac{1}{2}\,g_{D,[\bar 36]}^{(V)}\,{\rm tr}\,\big( \bar B_{[6]}\,i\,\gamma^\alpha\,\big\{ U_\alpha,\,U_\beta\big\}\,(D^\beta  B_{[\bar 3]})
- (D^\beta \bar B_{[6]})\,i\,\gamma^\alpha\,\big\{ U_\alpha,\,U_\beta\big\}\, B_{[\bar 3]}
+ {\rm h.c.} \big)
\nonumber\\
&& \qquad \;-\,  \frac{1}{2}\,g_{0,[66]}^{(V)}\left(\,{\rm tr}\,\big( \bar B_{[6]}\,i\,\gamma^\alpha\,(D^\beta B_{[6]})\big)\,{\rm tr}\,\big( U_\beta\,U_\alpha\big) + {\rm h.c.} \right)
\nonumber\\
&& \qquad \;-\, \frac{1}{4}\,g_{1,[66]}^{(V)}\,{\rm tr}\,\big( \bar B_{[6]}\,i\,\gamma^\alpha\,U_\beta\,(D^\beta B_{[6]})\, U^T_\alpha + \bar B_{[6]}\,i\,\gamma^\alpha\,U_\alpha\,(D^\beta B_{[6]})\, U^T_\beta + {\rm h.c.}\big)
\nonumber\\
&& \qquad \;-\, \frac{1}{2}\,g_{D,[66]}^{(V)}\,{\rm tr}\,\big( \bar B_{[6]}\,i\,\gamma^\alpha\,\big\{ U_\alpha,\,U_\beta\big\}\,(D^\beta  B_{[6]}) + {\rm h.c.}\big)
\nonumber\\
&& \qquad \;+\,\frac{1}{2}\, h_{0,[66]}^{(V)}\,{\rm tr}\,\big( \bar B_{[6]}^\mu\,g_{\mu \nu}\,i\,\gamma^\alpha\,(D^\beta B_{[6]}^\nu)\,{\rm tr}\,\big( U_\alpha\,U_\beta\big) + {\rm h.c.} \big)
\nonumber\\
&& \qquad \;+\,  \frac{1}{4}\,h_{1,[66]}^{(V)}\,{\rm tr}\,\big( \bar B_{[6]}^\mu\,g_{\mu \nu}\, i\,\gamma^\alpha\,U_\beta\,(D^\beta B_{[6]}^\nu)\, U^T_\alpha + \bar B_{[6]}^\mu\,g_{\mu \nu}\,i\,\gamma^\alpha\,U_\alpha\,(D^\beta B_{[6]}^\nu)\, U^T_\beta + {\rm h.c.}\big)
\nonumber\\
&& \qquad \;+\, \frac{1}{2}\,h_{2,[66]}^{(V)}\,{\rm tr}\,\big( \bar B^\mu_{[6]}\,g_{\mu \nu}\,i\,\gamma^\alpha\,\big\{ U_\alpha,\,U_\beta\big\}\,(D^\beta B_{[6]}^\nu ) + {\rm h.c.}\big)  \,,
\nonumber\\
%%%%%%%%%%%%%%%%%%%%%%%%%%%%%%%%%%%%%%%%%%
&&\mathscr{L}^{(A)} \!= f_{0,[66]}^{(A)}\,{\rm tr}\,\big( \bar B_{[6]}^\mu\,\gamma^\nu\,\gamma_5\,B_{[6]}\,{\rm tr}\,\big( U_\nu\,U_\mu\big) + {\rm h.c.} \big)
\nonumber\\
&& \qquad \;+\, f_{1,[66]}^{(A)}\,{\rm tr}\,\big( \bar B_{[6]}^\mu\,\gamma^\nu\,\gamma_5\,U_\nu\,B_{[6]}\,U^T_\mu
+ \bar B_{[6]}^\mu\,\gamma^\nu\,\gamma_5\,U_\mu\,B_{[6]}\,U^T_\nu + {\rm h.c.} \big)
\nonumber\\
&& \qquad \;+\,f_{D,[66]}^{(A)}\,{\rm tr}\,\big( \bar B_{[6]}^\mu\,\gamma^\nu\,\gamma_5\,\big\{ U_\mu,\,U_\nu\big\}\, B_{[6]} + {\rm h.c.}\big)
+ f_{F,[66]}^{(A)}\,{\rm tr}\,\big( \bar B_{[6]}^\mu\,\gamma^\nu\,\gamma_5\,\big[ U_\mu,\,U_\nu\big]\, B_{[6]} + {\rm h.c.}\big)
\nonumber\\
&& \qquad \;+\, f_{1,[\bar 36]}^{(A)}\,{\rm tr}\,\big( \bar B_{[6]}^\mu\,\gamma^\nu\,\gamma_5\,U_\nu\,B_{[\bar 3]}\,U^T_\mu
\textcolor{black}{-} \bar B_{[6]}^\mu\,\gamma^\nu\,\gamma_5\,U_\mu\,B_{[\bar 3]}\,U^T_\nu + {\rm h.c.} \big)
\nonumber\\
&& \qquad \;+\, f_{D,[\bar 36]}^{(A)}\,{\rm tr}\,\big( \bar B_{[6]}^\mu\,\gamma^\nu\,\gamma_5\,\big\{ U_\mu,\,U_\nu\big\}\, B_{[\bar 3]} + {\rm h.c.}\big)
+ f_{F,[\bar 36]}^{(A)}\,{\rm tr}\,\big( \bar B_{[6]}^\mu\,\gamma^\nu\,\gamma_5\,\big[ U_\mu,\,U_\nu\big]\, B_{[\bar 3]} + {\rm h.c.}\big) \,,
\nonumber\\
%%%%%%%%%%%%%%%%%%%%%%%%%%%%%%%%%%%%%%%%%%%%
&&\mathscr{L}^{(T)} \!= -g_{F,[\bar 3\bar 3]}^{(T)}\,{\rm tr}\,\big( \bar B_{[\bar 3]}\,i\,\sigma^{\alpha \beta}\,\big[ U_\alpha,\,U_\beta\big]\, B_{[\bar 3]}\big)
- g_{1,[\bar 36]}^{(T)}\,{\rm tr}\,\big( \bar B_{[6]}\,i\,\sigma^{\alpha \beta}\,U_\alpha\, B_{[\bar 3]}\,U^T_\beta + {\rm h.c.}\big)
\nonumber\\
&& \qquad \;-\, g_{F,[\bar 36]}^{(T)}\,{\rm tr}\,\big( \bar B_{[6]}\,i\,\sigma^{\alpha \beta}\,\big[ U_\alpha,\,U_\beta\big]\, B_{[\bar 3]} + {\rm h.c.}\big)
- g_{F,[66]}^{(T)}\,{\rm tr}\,\big( \bar B_{[6]}\,i\,\sigma^{\alpha \beta}\,\big[ U_\alpha,\,U_\beta\big]\, B_{[6]}\big)
\nonumber\\
&& \qquad \;+\, h_{F,[66]}^{(T)}\,{\rm tr}\,\big( \bar B^\mu_{[6]}\,g_{\mu \nu}\,i\,\sigma^{\alpha \beta }\,\big[ U_\alpha,\,U_\beta \big]\, B^\nu_{[6]}\big)\,,
\label{def-L2-detail}
\end{eqnarray}
where possible further terms are redundant owing to flavour identities or the  on-shell conditions
of spin-$\frac32$ fields with $\gamma_\mu\,B_{[6]}^\mu = 0$ and $\partial_\mu\,B_{[6]}^\mu = 0$. As compared to \cite{Lutz:2014jja} we further streamlined 
the notations and dropped the flavour redundant terms proportional to $g^{(S)}_{1,[\bar 3 6]}$ and $g^{(V)}_{1,[\bar 3 6]}$.

The symmetry conserving parameters contribute to the current-current correlation
function of two time-ordered axial-vector currents
\begin{eqnarray}
&& A^{\,ab}_{\mu \nu} (q) = i\,\int d^4 x \,e^{-i\,q\cdot x}  \,{\mathcal T}\,A^{(a)}_\mu (x)\,A^{(b)}_\nu(0) \,,
\label{def:axial-axial}
\end{eqnarray}
in the baryon states. 

The specific form of the matrix elements of the current-current correlation functions (\ref{def:scalar-scalar}) and  (\ref{def:axial-axial}) was already 
worked out in the previous work \cite{Lutz:2014jja}.  The matrix elements are detailed in the flavour SU(3) limit where the physical baryon states
are specified by the momentum $p$ and the flavour indices $i,j=1,2,3$.

\newpage

\section{Primer on large-$N_c$ operator analysis }

The low-energy constants recalled in (\ref{def-L4}) and (\ref{def-L2}) can be analyzed systematically in the $1/N_c$ expansion \cite{Dashen1994,Jenkins:1996de,Lutz:2010se,Lutz:2011fe,Lutz:2014jja}.
Leading order results have already been worked out in \cite{Lutz:2014jja}. Here we extend these results to the next accuracy level.

The large-$N_c$ operator expansion is performed in terms of a complete set of
static and color-neutral one-body operators that act on effective baryon states rather than the physical states \cite{Luty1994,Dashen1994,Jenkins:1996de,Lutz:2010se,Lutz:2014jja}.
In our case the physical and effective baryon states
\begin{eqnarray}
\ket{p,\, ij_\pm ,\,S,\,\chi }\,,  \qquad \qquad \qquad \qquad \roundket{ij_\pm,\,S,\, \chi}\,,
\label{def-states}
\end{eqnarray}
are specified by the momentum $p$ and the flavour indices $i,j,k=1,2,3$. The spin $S$ and the spin-polarization
are $\chi = 1,2$ for the spin one-half ($S=1/2$) and $\chi =1,\cdots ,4$ for the spin three-half states $(S=3/2)$. The
flavour sextet and the anti-triplet are discriminated by their symmetric (index $+$) and anti-symmetric (index $-$)
behaviour under the exchange of $i \leftrightarrow j$. At leading order in the $1/N_c$ expansion all considered baryon
states are mass degenerate.  The generic form of the operator expansion takes the form
\begin{eqnarray}
  \bra{\,\bar p,\,mn_\pm,\, \bar S,\,\bar\chi\,}\,{\mathcal O}_{QCD}\,
\ket{p,\,kl_\pm,\, S,\,\chi\,} = \sum_{n=0}^\infty\, c_n( \bar p, p)\,
\roundbra{\,mn_\pm,\,\bar S,\,\bar\chi \,} \,{\mathcal O}^{(n)}_{\rm static} \,
\roundket{\,kl_\pm,\,S,\,\chi\,} \,.
\label{def-largeN-expansion}
\end{eqnarray}
It is important to note that unlike the physical baryon states,
the effective baryon states do not depend on the momentum $p$. All dynamical information is moved into appropriate 
coefficient functions $c_n(\bar p, p)$. The contributions on the right-hand-side of (\ref{def-largeN-expansion}) can be sorted according to their relevance at large $N_c$. 

The effective baryon states $\roundket{ij_\pm, \chi}$  have a mean-field structure that can be
generated in terms of effective quark operators $q$ and $Q$ for the light and heavy species respectively. A corresponding
complete set of color-neutral one-body operators may be constructed in terms of the very same static quark operators
\begin{eqnarray}
&& \quarknumberoperator = q^\dagger ( \one  \otimes \one  \otimes \one )\,q \,, \qquad  \qquad \;\;\,
J_i = q^\dagger \Big(\frac{\sigma_i }{2} \otimes \one \otimes \one \Big)\, q \,,
\nonumber\\
&& T^a = q^\dagger \Big(\one \otimes \frac{\lambda_a}{2} \otimes \one \Big)\, q\, ,\qquad \quad \;\;
G^a_i = q^\dagger \Big( \frac{\sigma_i}{2} \otimes \frac{\lambda_a}{2} \otimes \one \Big)\, q\,,
\nonumber\\
&& \quarknumberoperator_{\!Q} = Q^\dagger ( \one  \otimes \one )\,Q \,, \qquad  \qquad \;\;\;\;\;\,
J_Q^{i} = Q^\dagger \Big(\frac{\sigma_i }{2} \otimes \one \Big)\, Q \,,
\label{def:one-body-operators}
\end{eqnarray}
with static operators $q=(u,d,s)^T$ and $Q= c$ of the up, down, strange and charm quarks. With $\lambda_a$ we
denote the Gell-Mann matrices supplemented with a singlet matrix $\lambda_0 = \sqrt{2/3}\,\one $.
Here we use a redundant notation with
\begin{eqnarray}
T^0 = \sqrt{\frac{1}{6}}\,\quarknumberoperator \,, \qquad \qquad G^0_i = \sqrt{\frac{1}{6}}\,J_i \,,
\label{def-redundant-operator}
\end{eqnarray}
which will turn useful when analyzing matrix elements of scalar currents.

In the sum of (\ref{def-largeN-expansion}) there are infinitely many terms one may write down. The static operators ${\mathcal O}_{\rm static}^{(n)}$ 
are finite products of the one-body operators $J_i\,, T^a$ and $G^a_i$. 
Terms that break the heavy-quark spin symmetry are exclusively caused by the heavy-spin operator
\begin{eqnarray}
J^i_Q \sim \frac{1}{M_Q} \,,
\label{def-spin-violation}
\end{eqnarray}
with the heavy-quark mass $M_Q$. In contrast the counting of $N_c$ factors is intricate since there is
a subtle balance of suppression and enhancement effects.
An $r$-body operator consisting of the $r$ products of any of the spin and flavour operators receives the
suppression factor $N_c^{-r}$. This is counteracted by enhancement factors for the
flavour and spin-flavour operators $T^a$ and $G^a_i$ that are
produced by taking baryon matrix elements at $N_c \neq 3$.
Altogether this leads to the effective scaling laws \cite{Dashen1994,Jenkins:1996de}
\begin{eqnarray}
J_i \sim \frac{1}{N_c} \,, \qquad \quad T^a \sim N^0_c \,, \qquad \quad G^a_i \sim N^0_c \,.
\label{effective-counting}
\end{eqnarray}
According to (\ref{effective-counting}) there is an infinite number of terms contributing at a given order in the
the $1/N_c$ expansion. Taking higher products of flavour and spin-flavour operators does not reduce the $N_c$
scaling power. A systematic $1/N_c$ expansion is made possible by a set of operator
identities \cite{Dashen1994,Jenkins:1996de,Lutz:2010se}, that allows a systematic summation of the infinite number
of relevant terms. This can be summarized into two reduction rules:
\begin{itemize}
\item All operator products in which two flavour indices are contracted using $\delta_{ab}$,
$f_{abc}$ or $d_{abc}$ or two spin indices on $G$'s are contracted using $\delta_{ij}$ or $\varepsilon_{ijk}$
can be eliminated.
\item All operator products in which two flavour indices are contracted using symmetric or antisymmetric
combinations of two different $d$ and/or $f$ symbols can be eliminated. The only exception to this rule is
the antisymmetric
combination $f_{acg}\,d_{bch}-f_{bcg}\,d_{ach}$.
\end{itemize}
As a consequence the infinite tower of spin-flavour operators truncates at any given order in the $1/N_c$ expansion.
We can now turn to the $1/N_c$ expansion of the baryon matrix elements of the QCD's axial-vector and scalar currents.
In application of the operator reduction rules, the baryon matrix elements of time-ordered products of the current
operators are expanded in powers of the effective one-body operators according to the counting rule (\ref{def-spin-violation}, \ref{effective-counting})
supplemented by the reduction rules. In contrast to Jenkins \cite{Jenkins:1996de} we consider the ratio $N_l/N_c= 1-1/N_c$ not as a suppression factor. The
strength of the spin-symmetry breaking terms we estimate with $1/M_Q \sim 1/N_c$. In the course of the construction of
the various structures, parity and time-reversal transformation properties are taken into account.

All what is needed in any practical application of the $1/N_c$ expansion is the action of any of the
one-body operators introduced in (\ref{def:one-body-operators}) on the effective mean-filed type baryon states
$\roundket{ij_\pm, \chi}$. In fact it suffices to provide results at the physical value $N_c =3$, for which a complete list was 
already generated in \cite{Lutz:2014jja}. We exemplify such results with
\begin{eqnarray}
 && J_Q^{k}\,|\,ij_+,\,{\textstyle{1\over 2}},\,\chi\,\big)
= -\frac 16\,\sigma^{(k)}_{\bar\chi\chi}\,|\,ij_+,\,{\textstyle{1\over 2}},\,\bar\chi\,\big)
+\frac{1}{\sqrt 3}\,\,S_{\bar\chi\chi}^{(k)}|\,ij_+,\,{\textstyle{3\over 2}},\,\bar\chi\,\big)
\,,\qquad 
\nonumber\\
&& J_Q^{k}\,|\,ij_+,\,{\textstyle{3\over 2}},\,\chi\,\big)
= \frac 12\,\big(\vec S\,\sigma^{(k)}\,\vec S^{\,\dagger}\,\big)_{\bar\chi\chi}\,|\,ij_+,\,{\textstyle{3\over 2}},\,\bar\chi\,\big)
+\textcolor{black}{\frac{ 1}{ \sqrt{3}}}\, S^{ (k) \dagger}_{\bar\chi\chi}\,|\,ij_{+},\,{\textstyle{1\over 2}},\,\bar \chi\,\big)\,,
\nonumber\\
&& J_Q^{k}\,|\,ij_-,\,{\textstyle{1\over 2}},\,\chi\,\big)
= \frac 12\,\sigma^{(k)}_{\bar\chi\chi}\,|\,ij_-,\,{\textstyle{1\over 2}},\,\bar\chi\,\big)\,,\qquad  \quad \;\;\,
\nonumber\\ \nonumber\\
&& J_k\,|\,ij_+,\,{\textstyle{1\over 2}},\,\chi\,\big)
= \frac 23\,\sigma_{\bar\chi\chi}^{(k)}\,|\,ij_+,\,{\textstyle{1\over 2}},\,\bar\chi\,\big)
-\,\frac{1}{\sqrt 3}\,\,S_{\bar\chi\chi}^{(k)}|\,ij_+,\,{\textstyle{3\over 2}},\,\bar\chi\,\big)\,,
\nonumber\\
&& J_k\,|\,ij_+,\,{\textstyle{3\over 2}},\,\chi\,\big)
= \big(\vec S\,\sigma^{(k)}\,\vec S^{\,\dagger}\,\big)_{\bar\chi\chi}\,|\,ij_+,\,{\textstyle{3\over 2}},\,\bar\chi\,\big)
-\,\frac{1}{\sqrt{3}}\,S_{\bar\chi\chi}^{(k)\,\dagger}\,|\,ij_+,\,{\textstyle{1\over 2}},\,\bar\chi\,\big)\,,
\nonumber\\
&& J_k\,|\,ij_-,\,{\textstyle{1\over 2}},\,\chi\,\big)
=  0\,,
\label{update-J}
\end{eqnarray}
where we apply the spin matrices $\sigma^{(k)}$ and $S^{(k)}$ in the convention as used in \cite{Lutz:2014jja}. Note that an error in the action of the heavy-spin operator $J_Q^k $ 
on the $|\,ij_+,\,{\textstyle{1\over 2}},\,\chi\,\big)$ states is corrected here in (\ref{update-J}). 
We affirm that now with (\ref{update-J}) the  relations 
\begin{eqnarray}
&& \big[J_i, \,J_j\big] = i\,\epsilon_{ijk}\,J_k\,,
\qquad \!\! \big[J^{i}_Q, \,J^{j}_Q\,\big] = i\,\epsilon_{ijk}\,J^{k}_Q\,, \qquad \;\;\big\{J^{i}_Q, \,J^{i}_Q\big\} = \frac{3}{2}\,\quarknumberoperator_{\!Q} \,, 
\nonumber\\
&& \big[J_Q^{i},\, J_j\big] =0 \,, \qquad \qquad 
\big[J_Q^{i},\, G^a_j\big] =0 \,, \qquad \qquad \quad \; \, \big[J_Q^{i},\, T^a\big] =0 \,,
\label{def-anti}
\end{eqnarray}
hold if matrix elements in the charmed baryon states as introduced in (\ref{def-states}) are taken. The latter were affected by the error made in \cite{Lutz:2014jja}.
Further corrections for matrix elements of the anti-commutator of two one-body operators are considered in the Appendix. 

\newpage

\section{Two scalar currents in charmed baryon matrix elements}

We turn to a derivation of large-$N_c$ sum rules for the chiral-symmetry breaking low-energy constants
introduced in (\ref{def-L4}). They contribute to the time-ordered product
of two scalar currents as evaluated in the baryon states.
At NLO in the $1/N_c$ expansion we find the relevance of 11 operators
\begin{eqnarray}
&& \bra{\,\bar p,\,mn_\pm,\, \bar S,\,\bar\chi\,}\,S^{ab}(q)\,
\ket{p,\,kl_\pm,\, S,\,\chi\,} =
\roundbra{\,mn_\pm,\,\bar S,\,\bar\chi \,} \,{\mathcal O}^{ab} \,
\roundket{\,kl_\pm,\,S,\,\chi\,}\,,
\nonumber\\
&& {\mathcal O}^{ab} =  \hat c_1\,\delta_{a0}\,\delta_{b0}\,{\mathbbm 1}
+ \,\hat c_2\,\delta_{ab}\,{\mathbbm 1}
+ \hat c_3\,\big( T_a\,\delta_{b0} + \delta_{a0}\,T_b \big) 
+ \hat c_4\,d_{abe}\,T_e +\hat c_5\,\{ T_a,T_b\}
\nonumber\\
&&  \qquad \! +\,  \hat c_6\,d_{abe}\,\{ J^i,G^i_e\}
 + \hat c_7\,\Big( \big\{ J^i\,,\,G^i_a \big\}\,\delta_{b0}
+ \delta_{a0}\,\big\{ J^i\,,\,G^i_b \big\}\Big)
\nonumber\\
&&  \qquad \! +\, \hat c_8 \,\Big( \big\{ J^i\,,\big\{T_a,\,\,G^i_b \big\}\big\} +\big\{ J^i\,,\big\{T_b,\,\,G^i_a \big\}\big\} \Big) 
\nonumber\\
&&  \qquad \! +\, 
\hat c_9\,d_{abe}\,\{ J_Q^i,G^i_e\}
 + \hat c_{10}\,\Big( \big\{ J_Q^i\,,\,G^i_a \big\}\,\delta_{b0}
+ \delta_{a0}\,\big\{ J_Q^i\,,\,G^i_b \big\}\Big)
\nonumber\\
&&  \qquad \! +\, \hat c_{11} \,\Big( \big\{ J_Q^i\,,\big\{T_a,\,\,G^i_b \big\}\big\} +\big\{ J_Q^i\,,\big\{T_b,\,\,G^i_a \big\}\big\} \Big)
+ \cdots   \,,
\label{def-SS-Nc}
\end{eqnarray}
where  only the first five operators that are required at LO were considered previously in \cite{Lutz:2014jja}.
We find six additional terms either involving the spin operators $J_i$ or $J_Q^i$. 
The three sums in (\ref{def-SS-Nc}) run over $e=1,... , 8$.

The operator truncation (\ref{def-SS-Nc}) can be matched to the tree-level Lagrangian (\ref{def-L4}). For this  the matrix elements of the operators in (\ref{def-SS-Nc}) are derived  
in the Appendix and  \cite{Lutz:2014jja}. Altogether we claim the identifications as detailed in Tab. \ref{tab:two-body-cc}.

At LO with $\hat c_{6-11} =0 $ there are $16-5=11$ sum-rules. No spin-symmetry breaking operator $J_Q î$ has to be considered at this accuracy level. 
In turn we recover the eight heavy-spin symmetry relations in (59) from \cite{Lutz:2014jja}. Additional relations arise from the large-$N_c$ considerations. 
We correct two misprints in (60) of \cite{Lutz:2014jja}. Altogether the following set of sum rules arises 
\begin{eqnarray}
&& c_{n,[66]} = e_{n,[66]}  \qquad {\rm for} \qquad n= 1,\cdots ,5 \,,\qquad
\nonumber\\
&& c_{n,[\bar 3 6]} = 0 \,\qquad  \quad \;\; {\rm for} \qquad n= 1,2\,,
\nonumber\\
%&& \textcolor{mono}{
%  c_{1,[\bar 3\bar 3]} = e_6\,, \qquad \qquad c_{2,[\bar 3\bar 3]} = e_7\,,\qquad \qquad
%   c_{3,[\bar 3\bar 3]} = e_8\,, \qquad \qquad
%   c_{4,[\bar 3\bar 3]} = e_9\,.}
% \nonumber\\
&& c_{1,[\bar 3\bar 3]} = c_{1,[66]} + \frac{1}{2}\,c_{5,[66]} \,, \qquad  \qquad\;
c_{2,[\bar 3\bar 3]} = c_{2,[66]}-\frac{1}{2}\,{c}_{5,[{6}{6}]}
 \,,
\nonumber\\
&& c_{3,[\bar 3\bar 3]} = c_{3,[66]}+2\,{c}_{5,[{6}{6}]}\,, \qquad  \qquad \;\;
 c_{4,[\bar 3\bar 3]} = c_{4,[66]}-2\,c_{5,[66]}  \,.
\label{c-LO-corrected}
\end{eqnarray}
At NLO all 11 operators in (\ref{def-SS-Nc}) turn relevant and we have $16-11=5$ sum rules
\begin{eqnarray}
 && \textcolor{black}{{c}_{1,[\bar{3}{6}]}
   = \frac{1}{\sqrt{3}\,}\,\big({c}_{3,[{6}{6}]}-{e}_{3,[{6}{6}]} \big) }
  \,,\qquad \qquad 
  {c}_{2,[{6}{6}]} = {e}_{2,[{6}{6}]}
  % ----------------------------------
  \,,\nonumber\\
  % ----------------------------------
&&\textcolor{black}{{c}_{2,[\bar{3}{6}]}
   = \,  \frac{1}{\sqrt{3}\, }\,\big( {c}_{4,[{6}{6}]}-{e}_{4,[{6}{6}]} \big) }
  % ----------------------------------
  \,,\nonumber\\
  % ----------------------------------
&& 3\,{c}_{1,[\bar{3}\bar{3}]}+{c}_{2,[\bar{3}\bar{3}]}+{c}_{4,[\bar{3}\bar{3}]}
   = \,
  3\,{c}_{1,[{6}{6}]}+{c}_{2,[{6}{6}]}+{c}_{4,[{6}{6}]}-{c}_{5,[{6}{6}]}
  % ----------------------------------
  \,,\nonumber\\
  % ----------------------------------
&& 3\,{e}_{1,[{6}{6}]}+{e}_{4,[{6}{6}]}-{e}_{5,[{6}{6}]}
   = \,
  3\,{c}_{1,[{6}{6}]}+{c}_{4,[{6}{6}]}-{c}_{5,[{6}{6}]}
  % ----------------------------------
  \,.
\end{eqnarray}
%%% ################################

\begin{table}[t]
\setlength{\tabcolsep}{2mm}
\renewcommand{\arraystretch}{1.2}
\begin{center}
\begin{tabular}{l|c||l|c}\hline
$c_{1,[66]} $              &$ \frac{2}{3}\,(3\,\hat{c}_{2}-\hat{c}_{4}-2\,\hat{c}_{6}) + \textcolor{black}{\frac{2}{3}\,\hat{c}_{9}}$                                     &$ e_{1,[66]} $            &$  \frac{1}{3}\,(6\,\hat{c}_{2}-2\,\hat{c}_{4}-4\,\hat{c}_{6}-\hat{c}_{9}) $\\
$c_{2,[66]} $              &$\frac{2}{3}\,\hat{c}_{1} $                                                                                    &$ e_{2,[66]} $            &$ \frac{2}{3}\,\hat{c}_{1} $\\
$c_{3,[66]} $              &$\sqrt{\frac{2}{3}}\,(2\,\hat{c}_{3}+4\,\hat{c}_{7}-\textcolor{black}{2 \,\hat{c}_{10}} ) $                               &$ e_{3,[66]} $            &$ \sqrt{\frac{2}{3}}\,(2\,\hat{c}_{3}+4\,\hat{c}_{7} +\textcolor{black}{\hat{c}_{10}} ) $\\
$c_{4,[66]} $              &$ 2\,\hat{c}_{4}+2\,\hat{c}_{5}+4\,\hat{c}_{6}+\textcolor{black}{8}\,\hat{c}_{8} -\textcolor{black}{2\,\hat{c}_{9} - 4\,\hat{c}_{11}}$     &$ e_{4,[66]} $            &$ 2\,\hat{c}_{4}+2\,\hat{c}_{5}+4\,\hat{c}_{6}+\textcolor{black}{8}\,\hat{c}_{8}+\hat{c}_{9} \textcolor{black}{\,+\,2\,\hat{c}_{11} } $\\
$c_{5,[66]} $              &$ 2\,\hat{c}_{5}+\textcolor{black}{8}\,\hat{c}_{8} \, \textcolor{black}{- 4\,\hat{c}_{11}}$                                               &$ e_{5,[66]} $            &$ 2\,\hat{c}_{5}+\textcolor{black}{8}\,\hat{c}_{8}  \textcolor{black}{\,+\,2\,\hat{c}_{11} }$\\
\hline
$c_{1,[\bar{3}\bar{3}]} $  &$ \frac{1}{3}\,(6\,\hat{c}_{2}-2\,\hat{c}_{4}+3\,\hat{c}_{5})$                                                 &$ c_{4,[\bar{3}\bar{3}]}$ &$  2\,(\hat{c}_{4}-\hat{c}_{5}) $\\
$c_{2,[\bar{3}\bar{3}]} $  &$ \frac{1}{3}\,(2\,\hat{c}_{1}-3\,\hat{c}_{5})$                                                                &$ c_{1,[\bar 36]}    $    &$ - \textcolor{black}{\sqrt{2}\,\hat{c}_{10}} $\\
$c_{3,[\bar{3}\bar{3}]} $  &$ 2\,\sqrt{\frac{2}{3}}\,\left(\hat{c}_{3}+\sqrt{6}\,\hat{c}_{5}\right)$                                       &$ c_{2,[\bar 36]} $       &$ -\textcolor{black}{\sqrt{3}\,(\hat{c}_{9}+2\,\hat{c}_{11} )} $\\ \hline

\end{tabular}
\caption{The symmetry breaking two-body counter terms as introduced in (\ref{def-L4}) are correlated by the large-$N_c$ operators (\ref{def-SS-Nc}). }
\label{tab:two-body-cc}
\end{center}
\end{table}

\clearpage

\section{Two axial currents in charmed baryon matrix elements}

We study the time-ordered product of two axial-vector currents.
The large-$N_c$ operator expansion was already worked out in \cite{Lutz:2010se} at leading order. Matrix elements in charmed baryons 
were derived in \cite{Lutz:2014jja}. Here we consider
and derive the implication for the chiral two-body interactions
introduced (\ref{def-L2}) at subleading orders in this expansion. At NLO there are 19 distinct operators to be considered
\begin{eqnarray}
&& \bra{\,\bar p,\,mn_\pm,\, \bar S,\,\bar\chi\,}\,A_{ij}^{ab}(q)\,
\ket{p,\,kl_\pm,\, S,\,\chi\,} =
\roundbra{\,mn_\pm,\,\bar S,\,\bar\chi \,} \,{\mathcal O}^{ab}_{ij} \,
\roundket{\,kl_\pm,\,S,\,\chi\,}\,,
\nonumber\\
&& \quad \,{\mathcal O}^{ab}_{ij} =  - \delta_{ij}\,\Big[
\, \hat g_1\,\delta_{ab}\,\quarknumberoperator+\hat g_2\,
d_{abc}\,T_c  + \hat g_3\,\{T_a,\,T_b\}  + \hat g_4\,d_{abc}\, \{J^k,\,G^k_c\} 
\nonumber\\
& & \qquad \qquad\quad + \,\hat g_5\,\Big( \{ J^k,\, \{T_{a},\,G_{b}^k\} \} + \{ J^k,\,\{G_{a}^k,\,T_{b} \} \Big)  
+  \hat g_6\,d_{abc}\, \{J_Q^k,\,G^k_c\} 
\nonumber\\
& & \qquad \qquad  \quad + \,\hat g_7\,\Big( \{ J_Q^k,\, \{T_{a},\,G_{b}^k\} \} + \{ J_Q^k,\,\{G_{a}^k,\,T_{b} \} \Big)  
\Big]
\nonumber\\
& & \qquad  \quad 
+ \,\hat g_8\,\Big( \{G_{a}^i,\,G_{b}^j\}  +\{G_{a}^j,\,G_{b}^i \} \Big)  +\hat g_{9}\,\Big( \big\{G_{a}^i,\,G_{b}^j\big\} - \big\{G_{a}^j,\,G_{b}^i \big\} \Big) 
\nonumber\\
&& \qquad \quad +\,\hat g_{20} \,d_{abc} \,\Big( \{G_{c}^i,\,J^j\}  + \{J^i,\,G_{c}^j\} \Big)
+\,\hat g_{21} \,d_{abc} \,\Big( \{G_{c}^i,\,J_Q^j\}  + \{J_Q^i,\,G_{c}^j\} \Big)
\nonumber\\
&& \qquad \quad +\, \epsilon_{ijk}\, f_{abc}\, \Big( \hat g_{10}\,G_c^k\,+ \hat g_{11}\,\{J^k,\,T_c\} + \hat g_{12}\,\{J_Q^k,\,T_c\}\Big)
\nonumber\\
&& \qquad \quad +\,\hat g_{22} \,i\,f_{abc} \,\Big( \{J^i,\,G_{c}^j\} -  \{G_{c}^i,\,J^j\} \Big)
+\,\hat g_{23} \,i\,f_{abc} \,\Big( \{J_Q^i,\,G_{c}^j\} -  \{G_{c}^i,\,J_Q^j\} \Big)
\nonumber\\
& & \qquad  \quad  + \,\frac{1}{4}\,(\bar p+p)_i \,(\bar p+p)_j \,\Big[ 
 \,\hat g_{13}\,\delta_{ab}\,\quarknumberoperator+ \hat g_{14}\,
d_{abc}\,T_c + \hat g_{15}\,\{T_a,\,T_b \} 
\nonumber\\
& & \qquad \qquad \quad+ \, \hat g_{16}\,d_{abc}\, \{J^k,\,G^k_c\} 
+ \hat g_{17}\,\Big( \{ J^k,\, \{T_{a},\,G_{b}^k\} \} + \{ J^k,\,\{G_{a}^k,\,T_{b} \} \Big)
\nonumber\\
& & \qquad \qquad \quad + \, \hat g_{18}\,d_{abc}\, \{J_Q^k,\,G^k_c\} 
+ \hat g_{19}\,\Big( \{ J_Q^k,\, \{T_{a},\,G_{b}^k\} \} + \{ J_Q^k,\,\{G_{a}^k,\,T_{b} \} \Big) \Big]
 + \cdots  \,,
\label{def-AA-Nc}
\end{eqnarray}
where we focus on the space components of the correlation function. In (\ref{def-AA-Nc})
we have $q= \bar p-p$ and $a,b =1, \cdots ,8$. In addition, we consider terms only that arise in the small-momentum
expansion and that are required for the desired matching with (\ref{def-L2}).
The dots in (\ref{def-AA-Nc}) represent additional terms that are further suppressed in the $1/N_c$ expansion or for small
3-momenta $p$ and $\bar p$. 

In the previous work \cite{Lutz:2014jja} only seven leading order operators  were considered. The ansatz of this work is reproduced with 
\begin{eqnarray}
 \hat g_1 = \frac{1}{3}\,\hat g_2\,, \qquad \quad  \hat g_{4-7} = 0 \,, \qquad \quad \hat  g_{11-12} = 0 \,,\qquad \quad  
  \hat g_{13} = \frac{1}{3}\,\hat g_{14}\,, \qquad \quad \hat g_{16-23} = 0 \,.
 \label{def-leading-AA-Nc}
\end{eqnarray}
An application of the results of our Appendix leads to the matching result as detailed in
Tab. \ref{tab:two-body}. 

\begin{table}[t]
\setlength{\tabcolsep}{1.mm}
\renewcommand{\arraystretch}{1.1}
\begin{center}
\begin{tabular}{l|c|l|c}\hline
  $ 
  {g}^{(S)}_{0,[\bar{3}\bar{3}]}$ & $2\,\hat{g}_{1}-\frac{2}{3}\,\hat{g}_{2}+\hat{g}_{3}+\frac{1}{2}\,\hat{g}_{8}
  $ & $\hat{g}^{(V)}_{0,[\bar{3}\bar{3}]} $ & $- \hat{g}^{(V)}_{1,[\bar{3}\bar{3}]} +2\,\hat{g}_{13}-\frac{2}{3}\,\hat{g}_{14}+\hat{g}_{15}
  $ \\ $ 
  {g}^{(T)}_{F,[\bar{3}\bar{3}]}$ & $-\hat{g}_{12}
  $ & & 
   \\ $
  {g}^{(S)}_{D,[\bar{3}\bar{3}]}$ & $\hat{g}_{2}-\hat{g}_{3}-\frac{3}{2}\,\hat{g}_{8}
  $ & $\hat{g}^{(V)}_{D,[\bar{3}\bar{3}]} $ & $2\,\hat{g}^{(V)}_{1,[\bar{3}\bar{3}]} + \hat{g}_{14}-\hat{g}_{15}
  $ \\ $
  {g}^{(S)}_{0,[{6}{6}]}$ & $2\,\hat{g}_{1}-\frac{2}{3}\,\hat{g}_{2} \,\textcolor{black}{-\,\frac{4}{3}\,\hat g_4 + \frac{2}{3}\,\hat g_6 + \frac{8}{9}\,(\hat g_{20} -\hat g_{21}/2) }
  $ & $\hat{g}^{(V)}_{0,[{6}{6}]}$ & $2\,\hat{g}_{13}- \frac{2}{3}\,\hat{g}_{14} \, \textcolor{black}{-\,\frac{4}{3}\,\hat g_{16} + \frac{2}{3}\,\hat g_{18}}
  $ \\ $
  {g}^{(S)}_{1,[{6}{6}]}$ & $2\,\hat{g}_{3}+8\,\hat{g}_{5}-4\,\hat{g}_{7}-\frac{1}{3}\,\hat{g}_{8}
  $ & $\hat{g}^{(V)}_{1,[{6}{6}]}$ & $2\,\hat{g}_{15}+8\,\hat{g}_{17}-4\,\hat{g}_{19}
  $ \\ $
  {g}^{(S)}_{D,[{6}{6}]}$ & $\hat g_{2-5}
  %$\hat{g}_{2}+\hat{g}_{3}+2\,\hat{g}_{4}+4\,\hat{g}_{5}
    -\hat{g}_{6}-2\,\hat{g}_{7}-\frac{1}{2}\,\hat{g}_{8} \textcolor{black}{- \frac{4}{3}\,(\hat g_{20}  -\hat g_{21} /2) }
  $ & $\hat{g}^{(V)}_{D,[{6}{6}]}$ & $\hat{g}_{14}+\hat{g}_{15}+2\,\hat{g}_{16}+4\,\hat{g}_{17}-\hat{g}_{18}-2\,\hat{g}_{19}
  $ \\ $
  {g}^{(T)}_{F,[\bar{3}{6}]}$ & $\frac{1}{2\,\sqrt{3}}\,(-\hat{g}_{9}+\hat{g}_{10})
   + \textcolor{black}{\frac{1}{\sqrt{3}}\, (\hat g_{22} - \hat g_{23}) }  
  $ & ${g}^{(T)}_{F,[{6}{6}]}$ & $\frac{1}{3}\,(\hat{g}_{9}-\hat{g}_{10}-4\,\hat{g}_{11}+\hat{g}_{12})
  $ \\ $
  {g}^{(T)}_{1,[\bar{3}{6}]}$ & $\frac{1}{\sqrt{3}}\,\hat{g}_{9}
  $ & ${f}^{(A)}_{1,[\bar{3}{6}]}$ & $-\hat{g}_{9}
  $ \\ &  & ${f}^{(A)}_{D,[\bar{3}{6}]}$ & $\textcolor{black}{\hat g_{21}}
  $ \\ $
  \hat{g}^{(V)}_{D,[\bar{3}{6}]}$ & $- \frac{\sqrt{3}}{2}\,(\hat{g}_{18}+2\,\hat{g}_{19})
  $ & ${f}^{(A)}_{F,[\bar{3}{6}]}$ & $-\hat{g}_{9}+\hat{g}_{10} 
  - \textcolor{black}{2\, (\hat g_{22} + \hat g_{23}/2) } 
  $ \\ $
  {g}^{(S)}_{D,[\bar{3}{6}]}$ & $-\frac{\sqrt{3}}{2}\,(\hat{g}_{6}+2\,\hat{g}_{7})
  + \textcolor{black}{\frac{1}{\sqrt{3}}\, \hat g_{21}}
  $ & $\hat{h}^{(V)}_{0,[{6}{6}]}$ & $2\,\hat{g}_{13}-\frac{2}{3}\,\hat{g}_{14}
   \textcolor{black}{\,-\,\frac{4}{3}\,\hat g_{16} -\frac{1}{3}\, \hat g_{18}}
  $ \\ $
  {h}^{(S)}_{0,[{6}{6}]}$ & $2\,\hat{g}_{1}-\frac{2}{3}\,\hat{g}_{2}\, \textcolor{black}{-\,\frac{4}{3}\,\hat g_4 - \frac{1}{3}\,\hat g_6 + \frac{4}{3}\,(\hat g_{20} +\hat g_{21}/2)}
  $ & $\hat{h}^{(V)}_{1,[{6}{6}]}$ & $2\,\hat{g}_{15}+8\,\hat{g}_{17}+2\,\hat{g}_{19}
  $ \\ $
  {h}^{(S)}_{1,[{6}{6}]}$ & $  \textcolor{black}{- \frac{4}{3}\,(\hat g_{20}  +\hat g_{21} ) }
  $ & $\hat{h}^{(V)}_{2,[{6}{6}]}$ & $\hat{g}_{14}+\hat{g}_{15}+2\,\hat{g}_{16}+4\,\hat{g}_{17}+\frac{1}{2}\,\hat{g}_{18}+ \hat{g}_{19}
  $ \\ $
  {h}^{(S)}_{2,[{6}{6}]}$ & $ \hat g_{2-5}
  %\hat{g}_{2}+\hat{g}_{3}+2\,\hat{g}_{4}+4\,\hat{g}_{5}
  +\frac{1}{2}\,\hat{g}_{6}+\hat{g}_{7}-\frac{1}{2}\,\hat{g}_{8} \textcolor{black}{- 2\,(\hat g_{20}  +\hat g_{21} /2) }
  $ & ${f}^{(A)}_{0,[{6}{6}]}$ & $ \textcolor{black}{- \frac{4}{3\,\sqrt{3}}\,(\hat g_{20}  -\hat g_{21}/2 ) }
  $ \\ $
  {h}^{(S)}_{3,[{6}{6}]}$ & $ \textcolor{black}{ 2\,(\hat g_{20}  +\hat g_{21} ) }
  $ & ${f}^{(A)}_{1,[{6}{6}]}$ & $\frac{1}{\sqrt{3}}\,\hat{g}_{8}
  $ \\ $
  {h}^{(S)}_{4,[{6}{6}]}$ & $2\,\hat{g}_{3}+8\,\hat{g}_{5}+2\,\hat{g}_{7}-\hat{g}_{8}
  $ & ${f}^{(A)}_{D,[{6}{6}]}$ & $ \textcolor{black}{ \frac{2}{\sqrt{3}}\,(\hat g_{20}  -\hat g_{21}/2 ) }
  $ \\ $
  {h}^{(S)}_{5,[{6}{6}]}$ & $ \textcolor{black}{2}\,\hat{g}_{8}
  $ & ${f}^{(A)}_{F,[{6}{6}]}$ & $\frac{1}{\sqrt{3}}\,(-\hat{g}_{9}+\hat{g}_{10}+4\,\hat{g}_{11}-4\,\hat{g}_{12})- \textcolor{black}{\sqrt{3}\,\hat g_{23}}
  $ \\ $
  {h}^{(T)}_{F,[{6}{6}]}$ & $\frac{1}{2}\,(\hat{g}_{9}-\hat{g}_{10}-4\,\hat{g}_{11}-2\,\hat{g}_{12})
  $ &  &  \\ \hline
\end{tabular}
\caption{The symmetry conserving two-body counter terms as introduced in (\ref{def-L2}) are correlated by the large-$N_c$ operators (\ref{def-AA-Nc}) with 
$\hat g^{(V)}_{n,[ab]} = 2\, g^{(V)}_{n,[ab]}/(M^{(1/2)}_{[a]}+ M^{(1/2)}_{[b]})$ and $\hat h^{(V)}_{n,[aa]} = h^{(V)}_{n,[aa]}/M^{(3/2)}_{[a]}$. We use also $\hat g_{2-5} = \hat{g}_{2}+\hat{g}_{3}+2\,\hat{g}_{4}+4\,\hat{g}_{5}$. 
In the flavour SU(3) chiral limit we 
use here three distinct charm baryon masses $M^{(J=1/2)}_{[\bar 3]}$, $M^{(J=1/2)}_{[6]}$ and $M^{(J=3/2)}_{[6]}$. Only the combinations
$g_{D,[\bar3\bar3]}^{(V)} -2\,g_{1,[\bar3\bar3]}^{(V)}$ and $g_{0,[\bar3\bar3]}^{(V)} +g_{1,[\bar3\bar3]}^{(V)}$ can be matched
at leading order in a non-relativistic expansion.}
\label{tab:two-body}
\end{center}
\end{table}

From the operator analysis (\ref{def-AA-Nc}), we obtain $33 -7=26$ sum rules. We do not reproduce all sum rules as considered first in \cite{Lutz:2014jja}.
In our analysis we unravel two misprint in (62) of \cite{Lutz:2014jja}. A complete list of relation reads
\begin{align}
  {g}^{(S)}_{D,[\bar{3}\bar{3}]}
  & = \,
  {h}^{(S)}_{2,[{6}{6}]} - {h}^{(S)}_{4,[{6}{6}]} - {h}^{(S)}_{5,[{6}{6}]}
  \,,\qquad
  {g}^{(S)}_{0,[\bar{3}\bar{3}]} = \frac{1}{2}\,\big( {h}^{(S)}_{4,[{6}{6}]} + {h}^{(S)}_{5,[{6}{6}]} \big)
  % -------------
  \,,\nonumber\\
  % -------------
  {g}^{(S)}_{0,[{6}{6}]}
  & = \,
  {g}^{(S)}_{D,[\bar{3}{6}]} = 0
  \,,\qquad\qquad\qquad
  {g}^{(S)}_{1,[{6}{6}]} = {h}^{(S)}_{4,[{6}{6}]} + \frac{1}{3}\,{h}^{(S)}_{5,[{6}{6}]}
  \,,\qquad
  {g}^{(S)}_{D,[{6}{6}]} = {h}^{(S)}_{2,[{6}{6}]}
  % -------------
  \,,\nonumber\\
  % -------------
  {h}^{(S)}_{0,[{6}{6}]}
  & = \,
  {h}^{(S)}_{1,[{6}{6}]} =
  {h}^{(S)}_{3,[{6}{6}]} = 0
  % ----------------------------------
  \,,\nonumber\\\nonumber\\
  % ----------------------------------
  \hat{g}^{(V)}_{0,[\bar{3}\bar{3}]}
  & = \,
  - \hat{g}^{(V)}_{1,[\bar{3}\bar{3}]} + \frac{1}{2}\,\hat{h}^{(V)}_{1,[{6}{6}]}
  \,,\qquad
  \hat{g}^{(V)}_{D,[\bar{3}\bar{3}]}
  =
  2\,\hat{g}^{(V)}_{1,[\bar{3}\bar{3}]} - \hat{h}^{(V)}_{1,[{6}{6}]} + \hat{h}^{(V)}_{2,[{6}{6}]}
  % -------------
  \,,\nonumber\\
  % -------------
  \hat{g}^{(V)}_{0,[{6}{6}]}
  & = \,
  \hat{g}^{(V)}_{D,[\bar{3}{6}]} =
  \hat{h}^{(V)}_{0,[{6}{6}]} = 0
  \,,\qquad
  \hat{g}^{(V)}_{1,[{6}{6}]} = \hat{h}^{(V)}_{1,[{6}{6}]}
  \,,\qquad
  \hat{g}^{(V)}_{D,[{6}{6}]} = \hat{h}^{(V)}_{2,[{6}{6}]}
  % ----------------------------------
  \,,\nonumber\\\nonumber\\
  % ----------------------------------
  {f}^{(A)}_{1,[\bar{3}{6}]}
  & = \,
  - \sqrt{3}\,{g}^{(T)}_{1,[\bar{3}{6}]}
  \,,\qquad
  {f}^{(A)}_{F,[\bar{3}{6}]} = -2\,{h}^{(T)}_{F,[{6}{6}]}
  \,,\qquad
  {f}^{(A)}_{D,[\bar{3}{6}]} = {f}^{(A)}_{0,[{6}{6}]} = {f}^{(A)}_{D,[{6}{6}]} = 0
  % -------------
  \,,\nonumber\\
  % -------------
  {f}^{(A)}_{1,[{6}{6}]}
  & = \,
  \frac{1}{2\,\sqrt{3}}\,{h}^{(S)}_{5,[{6}{6}]}
  \,,\qquad
  {f}^{(A)}_{F,[{6}{6}]} = -\frac{2}{\sqrt{3}}\,{h}^{(T)}_{F,[{6}{6}]}
  % ----------------------------------
  \,,\nonumber\\
  % ----------------------------------
  {g}^{(T)}_{F,[\bar{3}\bar{3}]}
  & = \,
  0
  \,,\qquad\qquad\qquad
  {g}^{(T)}_{F,[\bar{3}{6}]} = - \frac{1}{\sqrt{3}}\,{h}^{(T)}_{F,[{6}{6}]}
  \,,\qquad
  {g}^{(T)}_{F,[{6}{6}]} = \frac{2}{3}\,{h}^{(T)}_{F,[{6}{6}]}
  % ----------------------------------
  \,,
\label{large-Nc sumrules}
\end{align}
where the two identities in the fourth line of (\ref{large-Nc sumrules}) were not presented correctly in \cite{Lutz:2014jja}. 
We confirm the result of \cite{Lutz:2014jja} that the combination of the heavy spin-symmetry sum rules as summarized in (41) of \cite{Lutz:2014jja} 
with the large-$N_c$ sum rules (\ref{large-Nc sumrules})  does lead to one extra relation
\begin{eqnarray}
h^{(S)}_{5,[66]} = 0 \,.
\end{eqnarray}
As argued already in \cite{Lutz:2014jja} this does not contradict the systematics of the large-$N_c$ operator expansion. 
Though the operator analysis is not predicting such a feature, it can not exclude it.

We turn to the central result of our work. At NLO, we have derived $33-23=10$ novel sum rules
\begin{eqnarray}
 && {g}^{(S)}_{D,[\bar{3}{6}]}
   = \,
  \frac{1}{\sqrt{3}}\,\big({g}^{(S)}_{D,[{6}{6}]} - {h}^{(S)}_{2,[{6}{6}]} + \textcolor{black}{ \frac{1}{2}\, {h}^{(S)}_{1,[{6}{6}]} }\big)
  \,,\qquad \qquad 
  \qquad
 \textcolor{black}{  {h}^{(S)}_{3,[{6}{6}]} = - \frac{3}{2}\,{h}^{(S)}_{1,[{6}{6}]} }\,,
  % ----------------------------------
\nonumber\\
  % ----------------------------------
&&  \textcolor{black}{  3\,\hat{g}^{(V)}_{0,[{6}{6}]} +2\,\hat{g}^{(V)}_{D,[{6}{6}]} -\hat{g}^{(V)}_{1,[{6}{6}]} = 3\,\hat{h}^{(V)}_{0,[{6}{6}]} + 2\,\hat{h}^{(V)}_{2,[{6}{6}]} -\hat{h}^{(V)}_{1,[{6}{6}]}} \,,  \qquad \,\quad \,
   \hat{g}^{(V)}_{D,[\bar{3}{6}]}
   = \,
  \frac{1}{\sqrt{3}}\,\big(\hat{g}^{(V)}_{D,[{6}{6}]} - \hat{h}^{(V)}_{2,[{6}{6}]}\big) \,,
   % ----------------------------------
\nonumber\\
  % ----------------------------------
 &&  \textcolor{black}{ 3\,{g}^{(S)}_{0,[{6}{6}]} +2\,{g}^{(S)}_{D,[{6}{6}]} -{g}^{(S)}_{1,[{6}{6}]} = 3\,{h}^{(S)}_{0,[{6}{6}]} + 2\,{h}^{(S)}_{2,[{6}{6}]} -{h}^{(S)}_{4,[{6}{6}]} - \frac{1}{3}\,{h}^{(S)}_{5,[{6}{6}]} } \,,
  % ----------------------------------
\nonumber\\
  % ----------------------------------
&& {f}^{(A)}_{1,[\bar{3}{6}]}
   = \,
  - \sqrt{3}\,{g}^{(T)}_{1,[\bar{3}{6}]}
  \,,\qquad \qquad \qquad 
  {f}^{(A)}_{1,[{6}{6}]} =  \textcolor{black}{ \frac{1}{2\,\sqrt{3}} }\,{h}^{(S)}_{5,[{6}{6}]}
  \,,\qquad \qquad \qquad
{f}^{(A)}_{D,[{6}{6}]} = -\frac{3}{2}\, {f}^{(A)}_{0,[{6}{6}]} \,,
  % -------------
\nonumber\\
  % -------------
&&  \textcolor{black}{ {f}^{(A)}_{D,[\bar{3}{6}]}
   = \,
 - \frac{1}{\sqrt{3}}\,{f}^{(A)}_{D,[{6}{6}]} - \frac{1}{2}\,{h}^{(S)}_{1,[{6}{6}]} }\,,
  \qquad \qquad {g}^{(T)}_{F,[\bar{3}\bar{3}]}
   = \,
  - {g}^{(T)}_{F,[{6}{6}]} + \frac{2}{3}\,{h}^{(T)}_{F,[{6}{6}]} \,.
  % ----------------------------------
\end{eqnarray}
%%% ################################

\section{Summary}

In this work we considered the chiral Lagrangian for charmed baryons based on the flavour SU(3) symmetry. Current-current correlation functions in the baryon states were evaluated at tree-level and analyzed
in an expansion of $1/N_c$ at subleading orders. Sum rules for the symmetry breaking and symmetry conserving low-energy constants are systematically derived and 
presented. We correct various misprints found in previous works and establish novel sum rules that are valid at subleading orders in the $1/N_c$ expansion.   
Such parameter correlations are of crucial importance in  chiral extrapolation studies of the charmed baryon masses at N$^3$LO, but also they determine to a large extent the 
coupled-channel interaction of the charmed baryons with the set of Goldstone bosons. 

\vskip0.3cm
{\bfseries{Acknowledgments}}
\vskip0.3cm
We thank Xiao-Yu Guo for useful discussions. 
Y. Heo acknowledges partially support from Suranaree University of Technology, the Office of the Higher Education Commission under NRU project 
of Thailand (SUT-COE: High Energy Physics and Astrophysics) and  SUT-CHE-NRU (Grant No. FtR.11/2561).

%%%%%%%%%%%%%%%%%%%%%%%%%%%%%%%%%%%%%%%%%%%%%%%%%%%%%%%%%%%%%%%%%%%%%

\newpage
\section*{Appendix}

We consider matrix elements of the symmetric product of two one-body operators ${\mathcal O}$ in the
charmed baryon ground state at $N_c =3$. The generic notation
\begin{eqnarray}
&&\langle\, {\mathcal O} \,\rangle^{a\,b}_{\bar SS} \equiv \;
\roundbra{\,mn_a,\, \bar S, \,\bar\chi \,} \,{\mathcal O}\,\roundket{\,kl_{b},\,S,\,\chi\,}\,,
\end{eqnarray}
with $a,b = \pm$ will be applied. The results are expressed in terms of the flavour structures $\Lambda_{(kl)_+}^{(a),\,(rs)_\pm}$
and $\Lambda_{(kl)_-}^{(a),\,(rs)_\pm}$ and the spin structures $\sigma_i $ and $S_i$ as introduced
in \cite{Lutz:2014jja}. We correct and supplement the matrix elements that involve the spin operator $J_Q$ with 
\begin{eqnarray}
&& \langle\,\big\{\,J_Q^i,\,J_j\,\big\}\,\rangle^{--}_{\frac{1}{2}\frac{1}{2}}
=  0\,, \qquad \;\;\,
 \langle\,\big\{\,J_Q^i,\,J_Q^j\,\big\}\,\rangle^{--}_{\frac{1}{2}\frac{1}{2}}
= \frac 12\,\delta_{ij}\,\delta_{\bar\chi\,\chi}\,\delta_{(kl)_-}^{(mn)_-}\,,
\nonumber\\
&& \langle\,\big\{\,J_Q^i,\,T^a\,\big\}\,\rangle^{--}_{\frac{1}{2}\frac{1}{2}}
= \sigma_{\bar\chi\chi}^{(i)}\,\Lambda_{(kl)_-}^{(a),\,(mn)_-}\,, \qquad \;\,
 \langle\,\big\{\,J_Q^i,\,G^{a}_j\,\big\}\,\rangle^{--}_{\frac{1}{2}\frac{1}{2}}
= 0\,,
\nonumber\\
&& \langle\,\big\{\,J_Q^i,\,\big\{T^a\,, G^b_i\big\}\big\}\,\rangle^{--}_{\frac{1}{2}\frac{1}{2}}
= 0\,, \qquad \;\,
\nonumber\\ \nonumber\\
&& \langle\,\big\{\,J_Q^i,\,J_j\,\big\}\,\rangle^{+-}_{\frac{1}{2}\frac{1}{2}}
= 0\,, \qquad \;\;\,
 \langle\,\big\{\,J_Q^i,\,J_Q^j\,\big\}\,\rangle^{+-}_{\frac{1}{2}\frac{1}{2}}
=0\,,
\nonumber\\
&& \langle\,\big\{\,J_Q^i,\,T^a\,\big\}\,\rangle^{+-}_{\frac{1}{2}\frac{1}{2}}
= 0\,, \qquad \;\,
 \langle\,\big\{\,J_Q^i,\,G^{a}_j\,\big\}\,\rangle^{+-}_{\frac{1}{2}\frac{1}{2}}
= -\frac{1}{2\,\sqrt{3}}\,\Big(\delta_{ij} - i\,\epsilon_{ijk}\,\sigma_k \Big)_{\bar \chi \chi}\,\Lambda_{(kl)_-}^{(a),\,(mn)_+}\,,
\nonumber\\
&& \langle\,\big\{\,J_Q^i,\,\big\{T^a\,, G^b_i\big\}\big\}\,\rangle^{+-}_{\frac{1}{2}\frac{1}{2}} = - \frac{\sqrt{3}}{2}\,\delta_{\bar\chi\,\chi}\,\Big(
\Lambda_{(rs)_+}^{(a),\,(mn)_+}\,\Lambda_{(kl)_-}^{(b),\,(rs)_+} 
+ \Lambda_{(rs)_-}^{(b),\,(mn)_+}\,\Lambda_{(kl)_-}^{(a),\,(rs)_-} \Big)\,,
\nonumber\\ \nonumber\\
&& \langle\,\big\{\,J_Q^i,\,J_j\,\big\}\,\rangle^{++}_{\frac{1}{2}\frac{1}{2}}
=  - \,\frac{2}{3}\,\delta_{ij}\,\delta_{\bar\chi\,\chi}\,\delta_{(kl)_+}^{(mn)_+}\,, \qquad \;\;\,
 \langle\,\big\{\,J_Q^i,\,J_Q^j\,\big\}\,\rangle^{++}_{\frac{1}{2}\frac{1}{2}}
= \frac 12\,\delta_{ij}\,\delta_{\bar\chi\,\chi}\,\delta_{(kl)_+}^{(mn)_+}\,,
\nonumber\\
&& \langle\,\big\{\,J_Q^i,\,T^a\,\big\}\,\rangle^{++}_{\frac{1}{2}\frac{1}{2}}
=  - \frac{1}{3}\,\sigma_{\bar\chi\chi}^{(i)}\,\Lambda_{(kl)_+}^{(a),\,(mn)_+}\,, \qquad \;\,
 \langle\,\big\{\,J_Q^i,\,G^{a}_j\,\big\}\,\rangle^{++}_{\frac{1}{2}\frac{1}{2}}
=  - \,\frac{1}{3}\,\delta_{ij}\,\delta_{\bar\chi\,\chi}\,\Lambda_{(kl)_+}^{(a),\,(mn)_+}\,,
\nonumber\\
&& \langle\,\big\{\,J_Q^i,\,\big\{T^a\,, G^b_i\big\}\big\}\, \rangle^{++}_{\frac{1}{2}\frac{1}{2}} = 
-\delta_{\bar\chi\,\chi} \,\Big(
\Lambda_{(rs)_+}^{(a),\,(mn)_+}\,\Lambda_{(kl)_+}^{(b),\,(rs)_+} 
+ \Lambda_{(rs)_+}^{(b),\,(mn)_+}\,\Lambda_{(kl)_+}^{(a),\,(rs)_+} \Big)\,,
\nonumber\\ \nonumber\\
&& \langle\,\big\{\,J_Q^i,\,J_j\,\big\}\,\rangle^{+-}_{\frac{3}{2} \frac{1}{2}}\,
= 0\,, \qquad \;\;\,
 \langle\,\big\{\,J_Q^i,\,J_Q^j\,\big\}\,\rangle^{+-}_{\frac{3}{2} \frac{1}{2}}
= 0\,,
\nonumber\\
&& \langle\,\big\{\,J_Q^i,\,T^a\,\big\}\,\rangle^{+-}_{\frac{3}{2}\frac{1}{2}}
= 0\,, \qquad \;\;\,
 \langle\,\big\{\,J_Q^i,\,G^{a}_j\,\big\}\,\rangle^{+-}_{\frac{3}{2}\frac{1}{2}}
=  -\frac{1}{2}\,\Big(S_j\,\sigma_i \Big)_{\bar\chi\,\chi} \,\Lambda_{(kl)_-}^{(a),\,(mn)_+}\,,
\nonumber\\
&& \langle\,\big\{\,J_Q^i,\,\big\{T^a\,, G^b_i\big\}\big\}\,\rangle^{+-}_{\frac{3}{2}\frac{1}{2}} =  0\,,
\nonumber\\ \nonumber\\
&& \langle\,\big\{\,J_Q^i,\,J_j\,\big\}\,\rangle^{++}_{\frac{3}{2} \frac{1}{2} }
=  \frac{1}{\sqrt{3}} \,\Big(2\,S_i\,\sigma_j -S_j\,\sigma_i  \Big)_{\bar \chi \chi }
\,\delta_{(kl)_+}^{(mn)_+}\,, \quad
 \langle\,\big\{\,J_Q^i,\,J_Q^j\,\big\}\,\rangle^{++}_{\frac{3}{2} \frac{1}{2}}
=  0 \,,
\nonumber\\
&& \langle\,\big\{\,J_Q^i,\,T^a\,\big\}\,\rangle^{++}_{\frac{3}{2}\frac{1}{2} }
= \frac{2}{\sqrt{3} }\,S^{(i)}_{\bar\chi\chi}\,\Lambda_{(kl)_+}^{(a),\,(mn)_+}\,, \qquad \;
\nonumber\\
&& \langle\,\big\{\,J_Q^i,\,G^{a}_j\,\big\}\,\rangle^{++}_{\frac{3}{2}\frac{1}{2}}
= \frac{1}{2\,\sqrt{3} }\,\Big(2\,S_i\,\sigma_j -S_j\,\sigma_i  \Big)_{\bar\chi\,\chi}\,\Lambda_{(kl)_+}^{(a),\,(mn)_+}\,, \qquad 
\nonumber\\
&&  \langle\,\big\{\,J_Q^i,\,\big\{T^a\,, G^b_i\big\}\big\}\,\rangle^{++}_{\frac{3}{2} \frac{1}{2}} =0 \,,
\nonumber\\ \nonumber\\
&& \langle\,\big\{\,J_Q^i,\,J_j\,\big\}\,\rangle^{++}_{\frac{3}{2} \frac{3}{2}}
= \Big( \delta_{ij}- S_i\,S_j^{\dagger} - S_j\,S_i^{\dagger} 
\Big)_{\bar\chi\chi}\,\delta_{(kl)_+}^{(mn)_+}\,, \qquad
 \langle\,\big\{\,J_Q^i,\,J_Q^j\,\big\}\,\rangle^{++}_{\frac{3}{2} \frac{3}{2}}
=\frac{1}{2}\,\delta_{ij}\,\delta_{\bar\chi\,\chi}\,\delta_{(kl)_+}^{(mn)_+}\,,
\nonumber\\
&& \langle\,\big\{\,J_Q^i,\,T^a\,\big\}\,\rangle^{++}_{\frac{3}{2} \frac{3}{2}}
= \Big(\vec S\,\sigma_i\,\vec S^\dagger \Big)_{\bar\chi\chi}\,\Lambda_{(kl)_+}^{(a),\,(mn)_+}\,,
\nonumber\\
&& \langle\,\big\{\,J_Q^i,\,G^{a}_j\,\big\}\,\rangle^{++}_{\frac{3}{2} \frac{3}{2}}
= \frac{1}{2}\, \Big( \delta_{ij}
 -\,S_i\,S_j^{\dagger} - S_j\,S_i^{\dagger} 
\Big)_{\bar\chi\chi}\,\Lambda_{(kl)_+}^{(a),\,(mn)_+}\,,
\nonumber\\
&& \langle\,\big\{\,J_Q^i,\,\big\{T^a\,, G^b_i\big\}\big\}\,\rangle^{++}_{\frac{3}{2} \frac{3}{2}} = \frac{1}{2}\,\delta_{\bar\chi\,\chi} \,\Big(
\Lambda_{(rs)_+}^{(a),\,(mn)_+}\,\Lambda_{(kl)_+}^{(b),\,(rs)_+} 
+ \Lambda_{(rs)_+}^{(b),\,(mn)_+}\,\Lambda_{(kl)_+}^{(a),\,(rs)_+} \Big)\,,
\end{eqnarray}
where 
\begin{eqnarray}
2\,S_i\,\sigma_j -S_j\,\sigma_i =  \frac{1}{2}\,\Big( S_i\,\sigma_j + S_j\,\sigma_i \Big) -\frac{3}{2}\,i\,\epsilon_{ijk}\,S_k \,.
\end{eqnarray}

\newpage
\bibliography{1}

\end{document}